\documentclass[aps,prb,superscriptaddress,twocolumn,a4paper]{revtex4-1}

\usepackage{graphicx,amssymb,amsmath,amsfonts,bm,bbold}
\DeclareMathOperator{\sech}{sech}

\begin{document}

\title{Getting through to a qubit by magnetic solitons}

\author{Alessandro Cuccoli}
\affiliation{Dipartimento di Fisica e Astronomia, Universit\`a di Firenze,
	via G.~Sansone 1, I-50019 Sesto Fiorentino (FI), Italy}
\affiliation{Istituto Nazionale di Fisica Nucleare, Sezione di Firenze,
	via G.~Sansone 1, I-50019 Sesto Fiorentino (FI), Italy}

\author{Davide Nuzzi}
\affiliation{Dipartimento di Fisica e Astronomia, Universit\`a di Firenze,
	via G.~Sansone 1, I-50019 Sesto Fiorentino (FI), Italy}
\affiliation{Istituto Nazionale di Fisica Nucleare, Sezione di Firenze,
	via G.~Sansone 1, I-50019 Sesto Fiorentino (FI), Italy}

\author{Ruggero Vaia}
\affiliation{Istituto dei Sistemi Complessi, Consiglio Nazionale delle Ricerche,
	via Madonna del Piano 10, I-50019 Sesto Fiorentino (FI), Italy}
\affiliation{Istituto Nazionale di Fisica Nucleare, Sezione di Firenze,
	via G.~Sansone 1, I-50019 Sesto Fiorentino (FI), Italy}

\author{Paola Verrucchi}
\affiliation{Istituto dei Sistemi Complessi, Consiglio Nazionale delle Ricerche,
	via Madonna del Piano 10, I-50019 Sesto Fiorentino (FI), Italy}
\affiliation{Dipartimento di Fisica e Astronomia, Universit\`a di Firenze,
	via G.~Sansone 1, I-50019 Sesto Fiorentino (FI), Italy}
\affiliation{Istituto Nazionale di Fisica Nucleare, Sezione di Firenze,
	via G.~Sansone 1, I-50019 Sesto Fiorentino (FI), Italy}

\begin{abstract}
We propose a method for acting on the spin state of a spin-$\frac12$ localized particle, or {\it qubit}, by means of a magnetic signal effectively generated by the nearby transit of a magnetic soliton, there conveyed through a transmission line. We first introduce the specific magnetic soliton of which we will make use, and briefly review the properties that make it apt to represent a signal. We then show that a Heisenberg spin chain can serve as transmission line, and propose a method for injecting a soliton into the chain by acting just on one of its ends. We finally demonstrate that the resulting magnetic pulse can indeed cause, just passing by the spin-$\frac12$ localized particle embodying the qubit, a permanent change in its spin state, thus realizing the possibility of getting through to a single, localized qubit, and manipulate its state. A thorough analysis of how the overall dynamical system operates depending on the setting of its parameters demonstrates that fine tuning is not necessary as it exists an extended region in the parameters space that corresponds to an effective functioning. Moreover, we show that possible noise on the transmission line does not invalidate the scheme.
\end{abstract}

\maketitle

\section{Introduction}
\label{s.intro}

The ability of addressing, initializing, and possibly controlling one single qubit without spoiling its quantum features or disturbing other nearby qubits is a necessary prerequisite for putting a quantum device into operation. Depending on the specific device architecture, however, this can be a most challenging task, as it implies the opening of the qubit towards an environment that, in a way or another, embodies some macroscopic apparatus. One possibility for avoiding that this opening alter the fragile properties of the qubit is that of placing the apparatus at a distance, and use a transmission line for conveying a proper signal to the qubit itself. In particular, when the qubit is represented by a localized magnetic particle~\cite{TejadaCBHS2001,Engeletal2004,Gaebeletal2006,Mazeetal2008} it comes quite natural that the above signal be a time-dependent magnetic field, which nonetheless leaves the question open on how to realize the transmission line. To this purpose, we here propose the use of classical spin chains featuring soliton-like excitations, a choice suggested by these observations:
{\it i}) A soliton faithfully represents a signal in so far as it is a finite-energy excitation which is well localized in space at any given time, and can travel at fixed velocity with constant profile;
{\it ii}) Solitons are known to travel undisturbed along their medium, which allows us to put the apparatus that generates the pulse at great distance from the qubit and yet be sure that the signal will pass near its target undeformed;
{\it iii}) Solitons relative to the very same model can have different shapes and energies, which gives us the freedom of choosing the signal that best controls the qubit, without modifying the transmission line.

As for this latter component, we know that some classical fields defined on a one-dimensional space display solitonic excitations, whose renowned stability stems from the competition between linear and non-linear terms in the field's equations of motion (EoM). Based on the well established connection between classical vector-field theories and models of interacting spin-$S$ particles on discrete one-dimensional lattices~\cite{Auerbach1994}, one can expect soliton-like excitations to typify some spin chains~\cite{LongB1979,Fogedby1980,ElstnerM1989,KosevichIK1990,SchmidtSL2011,WoellertH2012}, which directly suggests that one such chain can serve for supporting a solitonic signal. Moreover, as extensively shown in the literature~\cite{KjemsS1978,BorsaPRT1983,PiniR1984,JensenMFHS1985,JohnsonW1985,MikeskaF1986} a renormalized classical approach~\cite{CTVV1991,CTVV1992,CGTVV1995,CTVV2000} is often appropriate for describing the actual behaviour of real compounds with $S>1/2$~\cite{BoucherRRRBS1980,RamirezW1982,MikeskaS1991}, which allows us to treat the signal's generation and propagation at a classical level.

The paper is structured as follows: In Sec.~\ref{s.solitons-as-signals} we introduce and characterize the soliton of which we will make use as non-linear excitation of a classical vector field. Based on the fact that such field is the continuum limit of a spin model on a one-dimensional lattice, namely the Heisenberg spin chain, in Sec.~\ref{s.Htransm} we numerically check that dynamical configurations corresponding to discrete versions of the above solitons can be solutions of the chain's EoM; in the same section we propose a method for injecting solitons by applying a time-dependent magnetic field to one end of the chain. In Sec.~\ref{s.qubit-dynamics} we thoroughly analyze the effects of the soliton's transit on the spin state of the spin-$\frac12$ localized particle embodying the qubit, considering both an ideal and an injected soliton, as well as the possibility of thermal noise along the chain. Comments upon possible experimental implementations of the scheme are put forward in Sec.~\ref{s.Conclusions}, with attention focused on the validity of the assumptions we have made in order to make our scheme function.

\section{Solitons as signals}
\label{s.solitons-as-signals}

\begin{figure}
	\hfill\includegraphics[width=0.48\textwidth]{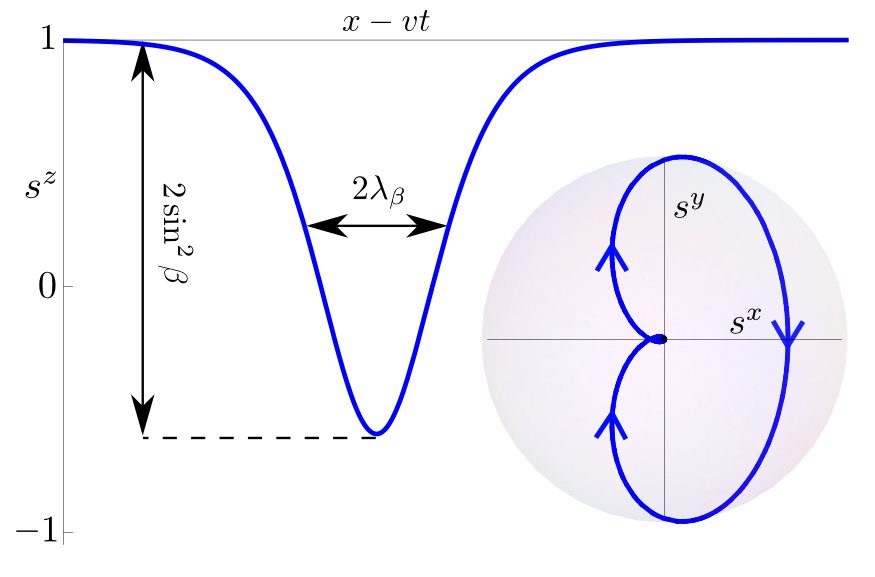}
	\caption{${\bm \Sigma^{(\beta)}}(x,t)$ for $\tan\!\beta\,{=}\,2$:  $s^z=1{-}2\sin^2\!\beta/\cosh^2\!\xi$ as a function of $x{-}vt$ and (inset) the corresponding trajectory of the in-plane components $s^x$ and $s^y$.}
	\label{f.soliton}
\end{figure}

Consider the classical vector-field in one spatial dimension, ${\bm s}(x,t)$, such that  $|{\bm s}(x,t)|\,{=}\,1$, with Hamiltonian density
\begin{equation}
 {\cal{H}}(x) = {\textstyle\frac12}{\cal J}{{\cal{S}}}^2 \,
                 \big[\partial_x\bm{s}(x,t)\big]^2
                +\gamma{H}{{\cal{S}}}\,\big[1-s^z(x,t)\big]~:
\label{e.Hcont}
\end{equation}
its EoM
\begin{equation}
 \partial_t\bm{s}(x,t)=
  \bm{s}(x,t) \times[{\cal J}{\cal{S}}~\partial_x^2\bm{s}(x,t)+
\gamma\bm{H}]~,
\label{e.sdotcont}
\end{equation}
with ${\bm H}=(0,0,H)$, have been shown~\cite{TjonW1977} to admit analytical solutions,  corresponding to localized excitations, stable under collisions~\cite{Takhtajan1977}, that travel at constant velocity. In polar coordinates these solutions  read
\begin{equation}
\bm{\Sigma}^{(\beta)}(x,t):
\left\{
\begin{aligned}
 \theta^{(\beta)}(x,t) &= 2\sin^{-1}(\sin\!\beta\,\sech\xi) ~,
\\
 \varphi^{(\beta)}(x,t) &= \varphi_0 + \cot\!\beta\,\xi
   +\tan^{-1}(\tan\!\beta\,\tanh\xi) ~,
\end{aligned}
\right.
\label{e.TWsol}
\end{equation}
where
\begin{equation}
\xi\equiv\,\frac{x{-}vt}{\lambda_{\beta}}=\frac{x}{\lambda_\beta}-\frac{t}{\tau_\beta} ~,
\end{equation}
and the parameter $\beta$ univocally characterizes each $v>0$ soliton, 
setting its characteristic 
\begin{eqnarray}
 &{\it amplitude}~:
 & 2\beta= 2\arccos \frac{v}{2\sqrt{{\cal J}{\cal S}\gamma H}}~,
\label{e.amplitude}\\
 &{\it length}~:
 & \lambda_\beta=\sqrt{\frac{{\cal J}{\cal{S}}}{\gamma{H}}}~\frac1{\sin\!\beta}~,
\label{e.sollength}\\
 &{\it energy}~:
 &\varepsilon_\beta= 8{\cal{S}}\sqrt{{\cal{J}}{\cal{S}}\gamma{H}}\,\sin\!\beta~,
\label{e.solen}\\
 &{\it time}~:~~~~~~~~~~
 &\tau_\beta=\frac1{\gamma{H}\sin2\beta}~.
\label{e.soltime}
\end{eqnarray}
A dynamical ($v>0$) soliton defined by Eqs.~\eqref{e.TWsol} will be hereafter referred to as "$\beta$-soliton". Notice that Eq.~\eqref{e.amplitude} sets a maximum  value for the velocity, $|v|<2\sqrt{{\cal JS}\gamma H }$, implying that the second term of the Hamiltonian \eqref{e.Hcont} must be finite in order for the model to support dynamical solitons. Once this condition is fulfilled, a $\beta$-soliton can be readily seen as a signal, i.e., a field's dynamical configuration with a distinctive trait that can be spotted, for time intervals of the order of $\tau_\beta$, in a spatial region of size $\lambda_\beta$ that moves with constant velocity in the one-dimensional space where the field is defined. An example of $\beta$-soliton is shown in Fig.~\ref{f.soliton}.

\section{Heisenberg chain as transmission line}
\label{s.Htransm}

Let us now consider a classical spin chain, i.e., a one-dimensional array of (spin-)vectors $\bm{S}_l\equiv{S}\,\bm{s}_l$, whose magnitude $S$ has the dimension of an action; the time dependence of the spin-variables will be hereafter understood, whenever possible. The Heisenberg chain is defined by the Hamiltonian
\begin{equation}
\begin{aligned}
 {\cal H}_{\rm chain} &=
 -{J} \sum_{l} \bm{S}_l{\cdot}\bm{S}_{l+1}
  -\gamma H\sum_l S_l^z
\\
  &=
 \frac{JS^2}2 \sum_l (\bm{s}_l{-}\bm{s}_{l+1})^2
   +\gamma{H}S\sum_l (1{-}s_l^z)+{\rm const}~,
\end{aligned}
\label{e.Hdiscr}
\end{equation}
where $J>0$ is a ferromagnetic nearest-neighbor coupling, 
$\bm{H}=(0,0,H)$ is an external magnetic field, and $\gamma$ is the 
gyromagnetic ratio.
In the (continuum) limit of vanishing 
lattice spacing $d\,{\to}\,0$ with finite $S/d$ and $Jd^3$, 
the model ~\eqref{e.Hdiscr} reproduces 
Eq.~\eqref{e.Hcont} with
\begin{equation}
{\cal{S}}\equiv\frac{S}{d}~~~{\rm and}~~~ {\cal{J}}\equiv Jd^3~.
\label{e.densities}
\end{equation}
The EoM for ${\bm s}_l$, obtained from the Poisson brackets $\{s_l^\alpha,s_j^\beta\} =S^{-1}\,\delta_{lj}\,\epsilon^{\alpha\beta\gamma}s_l^\gamma$, 
consistently have the same form of Eqs.~\eqref{e.sdotcont}, reading
\begin{equation}
 \partial_t\bm{s}_l
=\bm{s}_l\times\big[JS(\bm{s}_{l+1}{+}\bm{s}_{l-1})+\gamma\bm{H}\big]~.
\label{e.sdotdiscr}
\end{equation}

Despite the analogy, analytical soliton-like solutions of Eqs.~\eqref{e.sdotdiscr} are not known; however, as the continuum approximation does make sense whenever the relevant configurations vary slowly on the scale of the lattice spacing, we expect that, for $\lambda_\beta\gg d$, the discrete counterpart of a $\beta$-soliton, ${\bm \Sigma}_l^{(\beta)}(t)$ defined by Eqs.~\eqref{e.TWsol} with 
\begin{equation}
\xi=l\sqrt{\frac{\gamma H}{JS}}\sin\beta -\gamma Ht\sin 2\beta~,
\label{e.xidiscr}
\end{equation}
might still represent an excitation of the model \eqref{e.Hdiscr}. In fact, by numerically solving Eqs.~\eqref{e.sdotdiscr}, we have checked that the Heisenberg spin chain properly supports $\beta$-solitons whenever the Zeeman energy $\gamma H S$  is much smaller than the bond energy $JS^2$, as implied by $\lambda_\beta\gg d$ via Eqs.~\eqref{e.sollength} and \eqref{e.densities}. This result fits with the experimental observation, in quasi one-dimensional systems, of magnetic behaviours whose origin can be unequivocally ascribed to the presence of soliton-like excitations~\cite{RamirezW1982}. In what follows, we will therefore take that the time-dependent chain configuration $\big\{\bm{s}_l(t)=\bm{\Sigma}^{(\beta)}_l(t)\big\}$ is a solution of the discrete EoM \eqref{e.sdotdiscr}, embodying the signal we want to convey to the qubit, with the respective Heisenberg chain serving as transmission line.

\subsection{Injecting a soliton into the chain}
\label{ss.inject}

Let us now consider the problem of making one specific soliton $\bm{\Sigma}_l^{(\beta)}(t)$ to exist and run through the Heisenberg chain. In the process of accomplishing this goal, we first notice the following: Consider a finite (though long at will) chain, with $(2L+1)$ spins sitting on sites labelled from $-L$ to $+L$. Suppose ${\bm S}_{-L}(t)$ evolved as if a soliton were reaching it travelling from a fictitious, infinitely left-extended chain, $l<-L$: that soliton would continue travelling towards the region $l>-L$, i.e., it would be successfully injected into the chain, at least in the continuum limit. Therefore, enforcing
\begin{equation}
\bm{s}_{-L}(t)={\bm{\Sigma}}^{(\beta)}_{-L}(t)
\label{e.lefttail}
\end{equation}
as a boundary condition should result in the selection of the configuration corres\-ponding to ${\bm \Sigma}^{(\beta)}_l(t)$ , amongst all those that solve Eqs.~\eqref{e.sdotdiscr}. On the other hand it can be easily seen that Eqs.~\eqref{e.sdotdiscr} with condition \eqref{e.lefttail} enforced are the EoM of a Heisenberg chain with $-L<l\le L$, and an auxiliary magnetic field 
\begin{equation}
 \bm{b}^{(\beta)}(t)=\frac{JS}{\gamma}\bm{\Sigma}^{(\beta)}_{-L}(t)
\label{e.timedepfield}
\end{equation}
acting just on ${\bm S}_{-L+1}$. This suggests that, by applying the magnetic field~\eqref{e.timedepfield} to one end of the transmission line, it should be  possible to generate a soliton-like signal, that will then travel through to its target. The consistency of the above description has been checked, for different values of $\beta$, as follows. 

First, we have numerically solved Eqs.~\eqref{e.sdotdiscr} with the time-dependent constraint Eq.~\eqref{e.lefttail}, and $2Ld\,{\gg}\,\lambda_\beta$, by means of a second-order symplectic algorithm~\cite{KrechBL1998,Yoshida1990,ForestR1990,Suzuki1992,TsaiLL2005}. The chain has been initialized in the ferromagnetic configuration, $\{\bm{s}_l\,{=}\,\hat{\bm{z}}\}$, as well as in some possible thermal configurations. These have been determined, referring to the quadratic approximation of the Hamiltonian~\eqref{e.Hdiscr} which is diagonal in Fourier space with frequencies  $\omega_k=2JS(1{-}\cos{k})+\gamma H $, as inverse Fourier transforms of generated sets of independent variables $\{s^x_k,s^y_k\}$ with variances $\langle{s^x_ks^x_{-k}}\rangle=\langle{s^y_ks^y_{-k}}\rangle=k_{_{\rm{B}}}T/(S\omega_k)$.  The resulting configurations have the thermal correlators  $\big\langle(s^x_l{-}s^x_{l+1})^2\big\rangle=
\big\langle(s^y_l{-}s^y_{l+1})^2\big\rangle\simeq{\cal{T}}$ and  $\big\langle(s^x_l)^2\big\rangle=
\big\langle(s^y_l)^2\big\rangle\simeq{\cal{T}}\sqrt{JS/\gamma H}$, where ${\cal{T}}\equiv{k}_{_{\rm{B}}}T/JS^2$.

\begin{figure*}
	\includegraphics[width=0.85\textwidth]{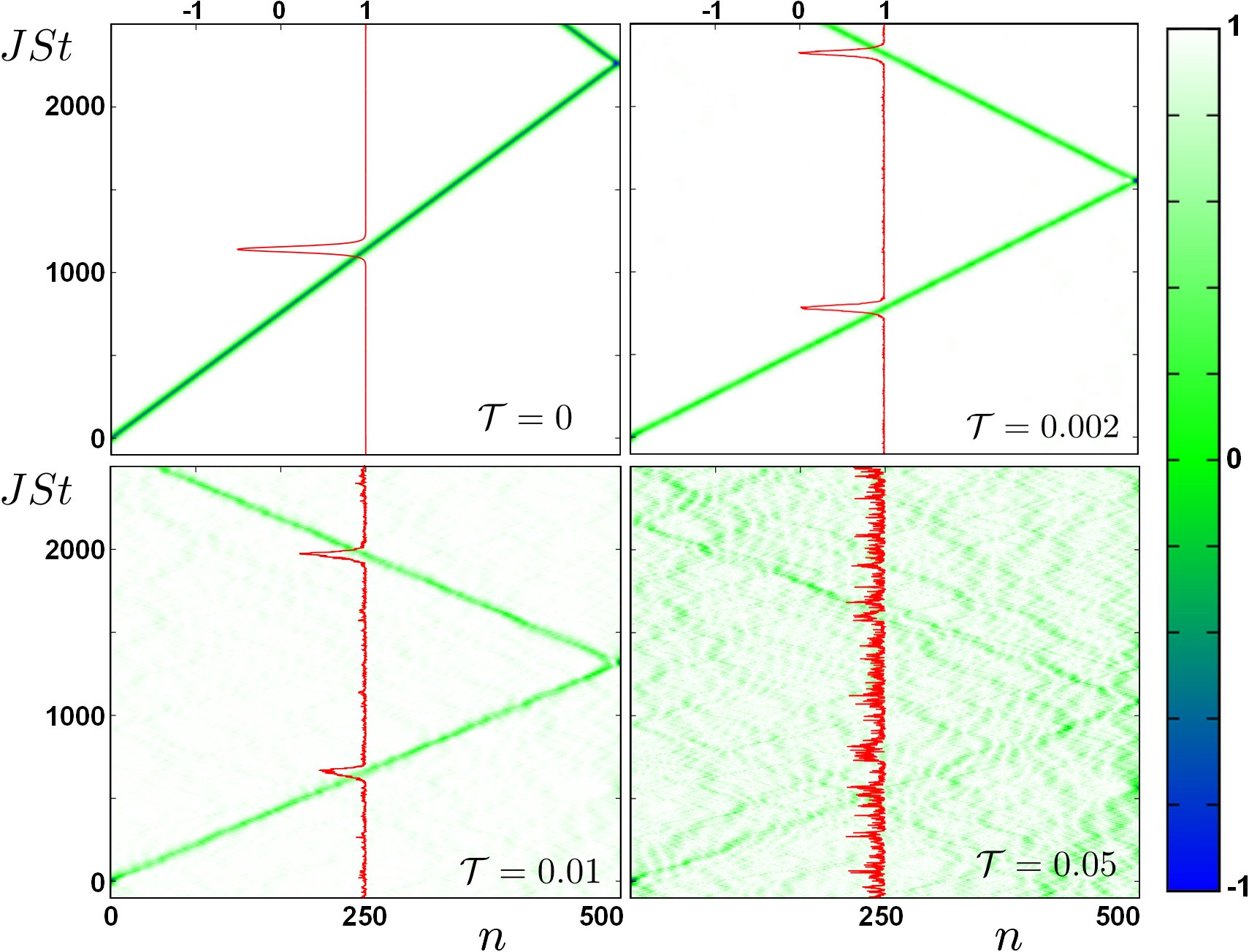}
	\caption{Samples of generated soliton-like excitations ${\bm \Gamma}_l(\beta;t)$ in a discrete chain of 500 spins; the parameters of the injected soliton ${\bm \Sigma}_l^{(\beta)}(t)$ are $\gamma H/JS=0.05$ and $\tan\!\beta\,{=}\,2$. Density plots are shown for the space-time evolution of $s_l^z(t)$ at zero (upper left panel) and finite temperature, indicated in the lower, right edge of each panel. The propagating soliton is reflected by the open boundary at site $n\,{=}\,500$. The thin, red line reports the time dependence of $s_l^z(t)$ at the site $n=250$. Thermal noise makes the generated  soliton-like excitation broader and faster.}
	\label{f.gensol}
\end{figure*}

We have then analyzed the resulting solutions and found that when the field ${\bm b}^{(\beta)}(t)$ is applied to ${\bm s}_{-L+1}$, i.e., after the injection of ${\bm \Sigma}^{(\beta)}_l(t)$, dynamical configurations ${\bm \Gamma}_l(\beta;t)$, identifiable as soliton-like, actually appear into the chain. In order to better characterize these configurations, we have numerically measured their velocity $v'$ and, assuming the validity of Eq.~\eqref{e.amplitude}, we have obtained values for the respective amplitudes, $2\beta'=2\arccos[v'/(JS\gamma H)]$. These values have been found to agree with those independently determined by fitting $\Gamma_l(\beta;t)$ with Eqs.~\eqref{e.TWsol}, for all values of $\beta$ considered. Moreover, by monitoring the chain's energy throughout the numerical integration, we have calculated the total work made by the forcing term, and found it very close to $\varepsilon_{\beta'}$, meaning that the work done on the chain does actually correspond to the energy needed to generate a soliton ${\bm\Sigma}^{(\beta')}_l(t)$. Summarizing, the above analysis confirms that:
\begin{itemize}
 \item by applying the field ${\bm b}^{(\beta)}(t)$ defined in Eq.~\eqref{e.timedepfield} to the left tail of the chain,
 \item a dynamical configuration ${\bm \Gamma}_l(\beta;t)$ is generated inside the chain itself, 
 \item with the essential features of a soliton ${\bm\Sigma}_l^{(\beta')}(t)$.
\end{itemize}
The above picture is also confirmed for ${\cal T}>0$, and even for rather narrow injected solitons ($\lambda_\beta\simeq 5d$). In Fig.~\ref{f.gensol} we show the colour-density plot of $\Gamma^z_l(\beta;t)$ as a function of $l$ and $JSt$, for different ${\cal T}$. The strong resilience of the generated signal is evident: even when fully embedded and barely recognizable within the thermal noise, as in the last panel, its time propagation along the chain can still be easily followed.

\section{Qubit dynamics}
\label{s.qubit-dynamics}

\begin{figure}
	\includegraphics[width=0.48\textwidth]{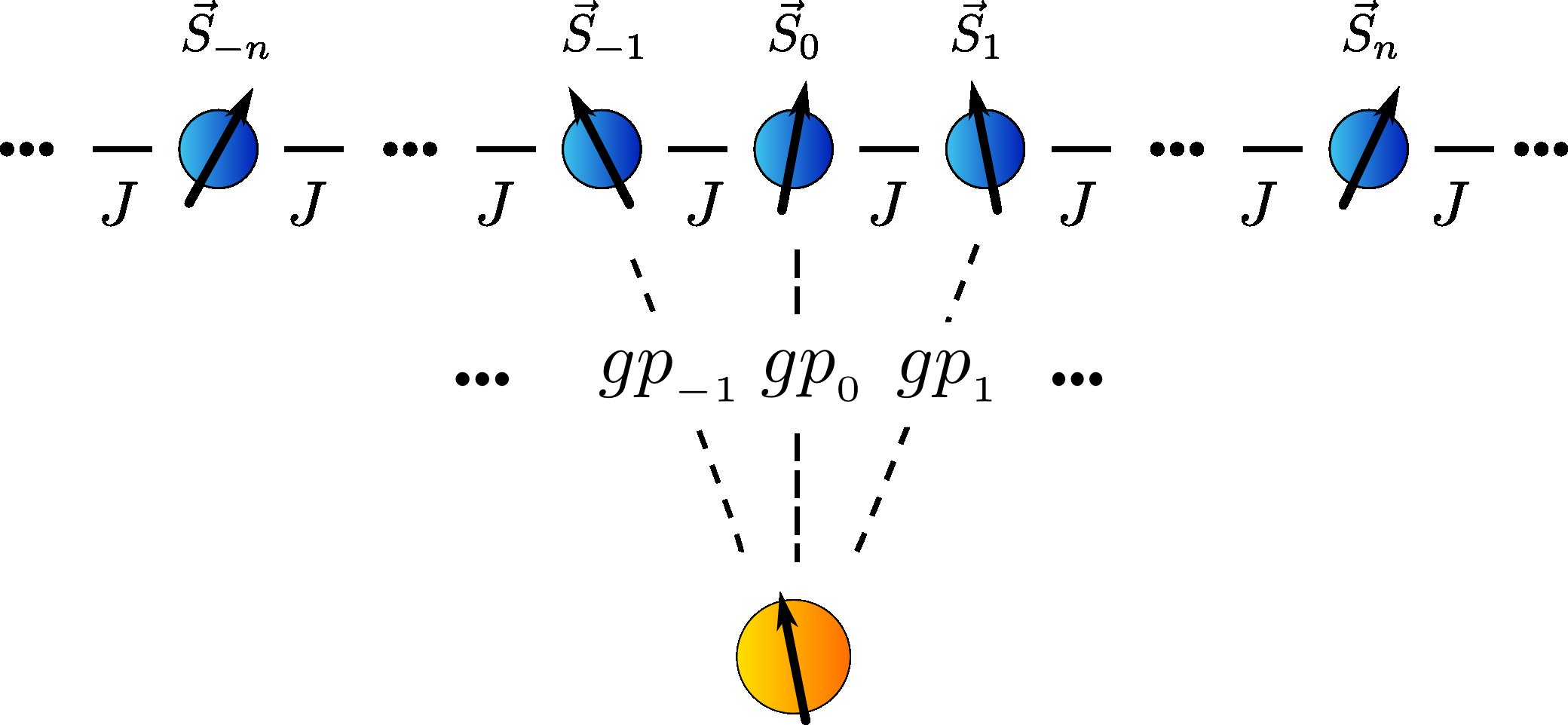}
	\caption{The qubit interacts with a bunch of moments of the classical spin-chain, with couplings $j_l=g\,p_l$; a constant uniform field is applied to the overall system.}
	\label{f.qubit-chain}
\end{figure}

A qubit is a physical system that can be described by the spin-$\frac12$ operator $\frac\hbar{2}\hat{\bm{\sigma}}$ represented by the Pauli matrices  $\hat{\bm{\sigma}}=(\hat{\sigma}^x,\hat{\sigma}^y,\hat{\sigma}^z)$; its  state $\hat\rho(t)$ in terms of the Bloch vector $\bm{n}(t)\equiv{\rm{Tr}}\big[\hat\rho(t)\,\bm{\hat\sigma}\big]$, reads
\begin{equation}
 \hat\rho(t) = 
\frac12\big[\mathbb{1}+\bm{\hat\sigma}{\cdot}\bm{n}(t)\big]~.
%  = \frac12 \begin{pmatrix}
%  1{+}n^z    & n^x{-}in^y \\
%  n^x{+}in^y & 1{-}n^z \end{pmatrix} ~.
\label{e.rhoqubit}
\end{equation}
In our scheme, the qubit is realized by a localized spin-$\frac12$ particle, sitting near the site of the chain labelled by the index ``0''. We assume that the way the qubit feels the presence of the chain's magnetic moments, as depicted in Fig.~\ref{f.qubit-chain}, can be generally described as a Zeeman interaction with an effective magnetic field proportional to
\begin{equation}
 \tilde{\bm s}(t)\equiv\sum_j p_j{\bm s}_j(t)~,
\label{e.stildedisc}
\end{equation}
where ${\bm s}_j$ are the unit vectors entering Eq.~\eqref{e.Hdiscr}, and $p_j$ is expected to decrease rapidly with $|j|$. In fact, the detailed dependence of $p_j$ on $j$ is not relevant, particularly if, as in the present scheme, the time dependence of the magnetic moments is primarily due to the transit of a signal whose length is of a finite number of lattice spacings. Therefore, we can safely choose a Gaussian dependence $p_j=A\exp(-j^2/2\alpha^2)$~, where $A$ is such that $\sum_j p_j = 1$~, and the standard deviation $\alpha$ characterizes the interaction range, in units of $d$.
  
From Secs.~\ref{s.solitons-as-signals} and~\ref{s.Htransm} we have learned that the presence of a constant and homogeneous magnetic field $\bm{H}$ is necessary for the Heisenberg spin chain to support solitons with finite velocity: therefore, we take $\bm{H}\neq 0$ and identify its direction with the quantization axis used for encoding the qubit states into the spin degree of freedom of the spin-$\frac12$ localized particle. The qubit's Hamiltonian thus reads
\begin{equation}
  \hat{\cal{H}}_{\rm qubit} =
  -\big[\gamma_\sigma H \,\hat{\bm{z}}+g\,S\,\tilde{\bm s}(t)\big]
  \cdot\frac{\hbar\bm{\hat\sigma}}2  ~,
\label{e.Hqubit}
\end{equation}
where  $\gamma_\sigma$ is the gyromagnetic ratio of the particle realizing the qubit, and $g$ is an overall coupling constant. The corresponding evolution of the qubit's Bloch vector is ruled by the equation
\begin{equation}
 \partial_\tau\bm{n} = 
 \bm{n}\times\big[\delta\,\hat{\bm{z}}+\mu\,\tilde{\bm{s}}(\tau)\big] ~,
\label{e.dtaua}
\end{equation}
where $\hat{\bm{z}}=(0,0,1)$, and $\tau\equiv\gamma{H}\,t$ is the (chain's) dimensionless time that will be hereafter used. The two dimensionless parameters
\begin{equation}
  \delta=\frac{\gamma_\sigma}{\gamma}
  ~,~~~~~~
  \mu \equiv \frac{gS}{\gamma{H}}
\label{e.deltamu}
\end{equation}
characterize the qubit's interactions with the external and the effective field, ${\bm H}$ and $\tilde{\bm s}(t)$, respectively. Notice that, despite the chain parameter $\gamma$ does not appear in the qubit's Hamiltonian~\eqref{e.Hqubit}, it does enter the EoM for the qubit's Bloch vector via the definition of the dimensionless time $\tau$; in fact, the relevant time scale of the overall dynamics is exclusively set by the chain Hamiltonian~\eqref{e.Hdiscr}, a statement based on the implicit assumption that the presence of the qubit has no effect (no `back-action') on the chain itself. We will further comment upon this assumption in Sec.\ref{s.Conclusions}.

Suppose now that a magnetic signal in the form of ${\bm\Gamma}_l(\beta;\tau)$ run through the chain (${\cal T}=0$). In the early stage of the process, at a time when the soliton is still far from the site $0$, it is $\tilde{\bm{s}}(\tau)\,{\propto}\,\hat{\bm{z}}$ and the qubit Bloch vector undergoes a uniform precession around $\hat{\bm{z}}$, unless it is not initially aligned along the $z$-axis itself. In order to isolate the qubit evolution exclusively due to the soliton transit, it is therefore convenient to choose $\bm{n}(\tau_{\rm{i}}){=}\hat{\bm{z}}$, with $\tau_{\rm{i}}$ the earlier time when ${\bm s}_{-L}\neq\hat{\bm z}$. Notice that this does not imply adding a previous single-qubit manipulation step in the overall scheme, but rather preparing the whole system in a globally aligned state, which is readily obtained as ${\bm H}\neq{0}$.

Consider now a time $\tau_{\rm f}$ during the final stage of the process, i.e. after the soliton has travelled along the chain far beyond the qubit: the qubit's Bloch vector, set in motion by the soliton's transit, can {\it 1)} align back to $\hat{\bm z}$, {\it 2)} tilt-up and hence precess around $\hat{\bm z}$, or {\it 3)} perfectly flip and anti-align along $-\hat{\bm z}$. Situations {\it 2)} and {\it 3)} are those in which we are most interested, as they represent the possibility of permanently modifying the qubit state, which is in fact the final goal of our scheme. In order to analyze the conditions under which they are obtained, one must numerically integrate Eqs.~\eqref{e.dtaua} with the effective field $\tilde{\bm s}(\tau)$ as from Eq.~\eqref{e.stildedisc} with
\begin{equation}
{\bm s}_j(\tau)=
{\bm \Gamma}_j(\beta;\tau)~.
\label{e.spreadsoliton}
\end{equation}
For the sake of clarity, in what follows we will specifically 
concentrate on the case 
when the qubit's response to the signal consists in a permanent 
flipping.

\subsection{qubit flipped by an ideal soliton}
\label{ss.ideal}

\begin{figure*}
 \includegraphics[width=0.8\textwidth]{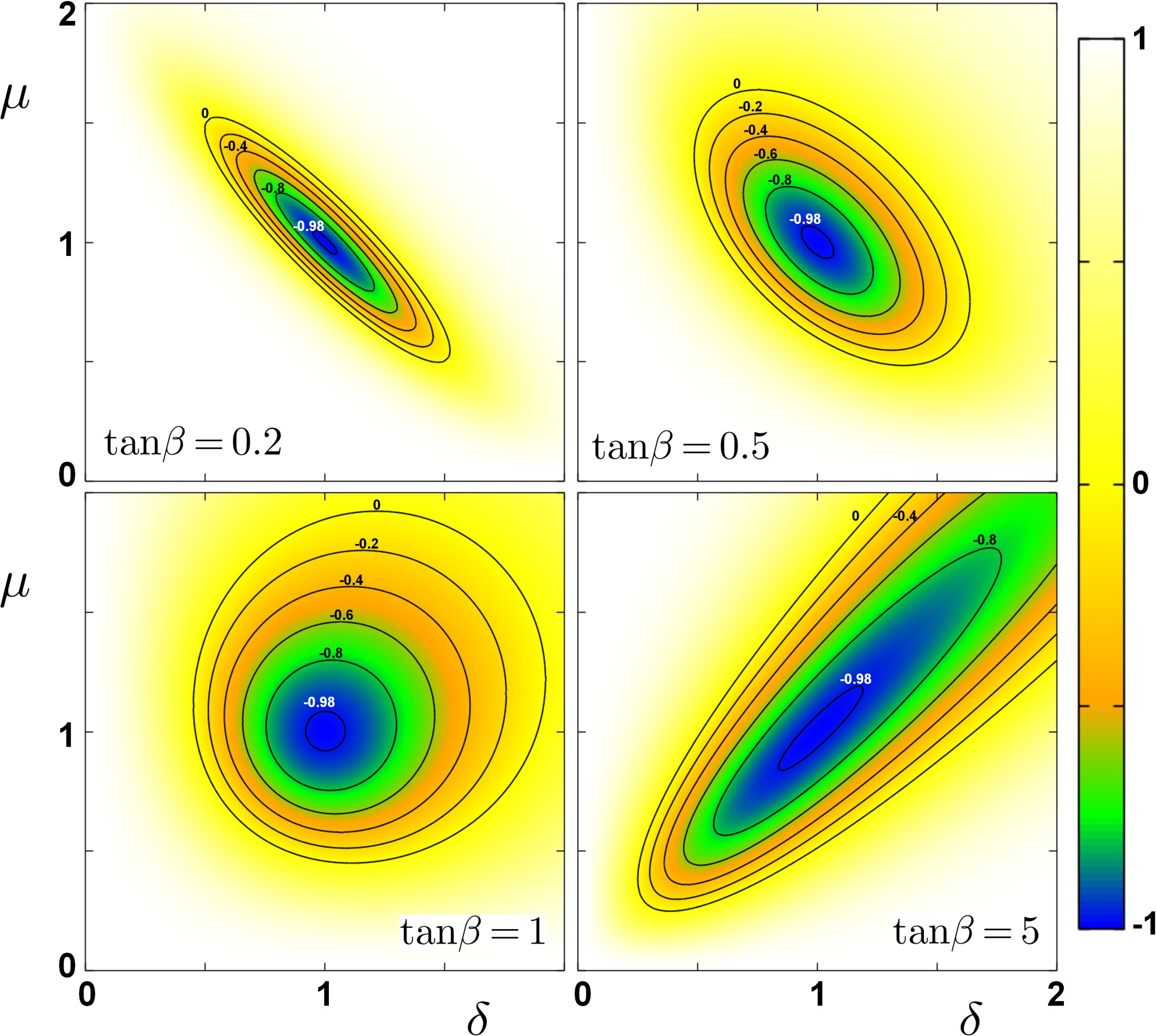}
 \caption{Contour plots of $n^z(\tau_{\rm{f}})$ as a function of $\delta$ and $\mu$, for $\alpha\,{=}\,0$. In each panel the magnetic signal acting on the qubit is that produced by an ideal $\beta$-soliton, ${\bm \Sigma}_l^{(\beta)}(t)$.}
\label{f.az_contour}
\end{figure*}

Let us first take ${\bm \Gamma}_l(\beta;\tau)={\bm \Sigma}_l^{(\beta)}(\tau)$ and the chain initially prepared in the ferromagnetic state (${\cal T}=0$). When $\alpha\,{=}\,0$ we know~\cite{CNVV2014a} that whenever $\delta\,{=}\,0$ the qubit always goes back to its initial state: therefore, in order to obtain a permanent flipping, the physical object embodying the qubit must have a finite gyromagnetic ratio. As studying ${\bm n}^z(\tau_{\rm f})$ suffices to distinguish the above situations {\it 1)}, {\it 2)}, and {\it 3)}, in Fig.~\ref{f.az_contour} we plot $n^z(\tau_{\rm f})$ in the plane $(\delta,\mu)$: when $\delta\,{=}\,\mu\,{=}\,1$ the flipping is complete, while the change in $n^z(\tau_{\rm f})$ decreases monotonically when getting far from this point. Remarkably, for $\delta\,{=}\,\mu\,{=}\,1$ there is no dependence on $\beta$: the qubit is flipped whatever the amplitude of the signal running through the chain. An additional feature, numerically observed and clearly seen in Fig.~\ref{f.az_contour}, is that $n^z(\tau_{\rm f})$ is symmetric in the exchange $\delta\leftrightarrow\mu$, even though the evolution of the qubit may be different in the two cases. The most relevant feature displayed by Fig.~\ref{f.az_contour}, however, is the presence of a region where almost complete flipping occurs: this means that fine-tuning is not necessary and if $\delta$ is difficult to alter one can still act on $\mu$, or viceversa, depending on the specific physical realization of the scheme.

\begin{figure}
	\includegraphics[width=0.48\textwidth]{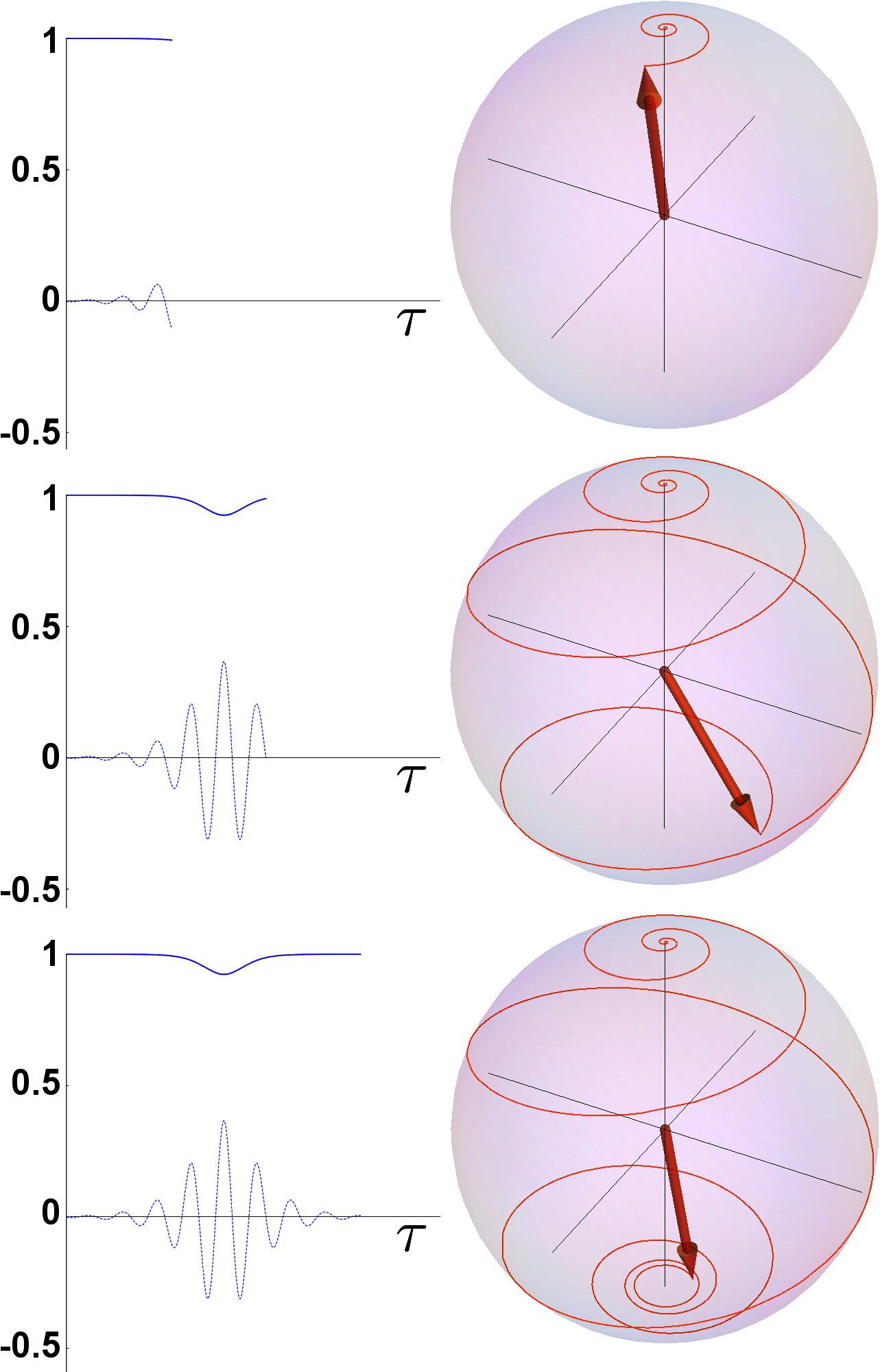}
	\caption{Time lapse of qubit's evolution (represented by $\bm{n}(\tau)$) while a $\beta$-soliton propagates along the chain. The graphs on the left side display ${\tilde{s}}^z(\tau)$ (full lines) and ${\tilde{s}}^x(\tau)$ (dashed lines), i.e., the components of the effective field acting on the qubit as a consequence of the soliton's transit, drawn up to the same time $\tau$. The parameter values are: $h\,{=}\,0.01$, $\tan\!\beta\,{=}\,0.2$, $\mu\,{=}\,1$, $\delta\,{=}\,1$, and $\alpha\,{=}\,3$.}
	\label{f.lapseopt}
\end{figure}

When a finite interaction range ($\alpha\,{\neq}\,0$) is considered, as in the case shown in Fig.~\ref{f.lapseopt}, the qubit's dynamics is qualitatively similar to that observed for $\alpha\,{=}\,0$~\cite{CNVV2014b}, but the value of $n^z(\tau_{\rm f})$ is found quite sensitive to $\alpha$ itself. However, an almost complete flipping, even better than that observed in Fig.~\ref{f.lapseopt}, can be obtained by further adjusting the available parameters. To this respect, notice that the ratio $h\,{=}\,{\gamma H}/{JS}$ is a relevant quantity, when $\alpha\neq 0$, as it contributes to set the length scale of the soliton: For example, $h\,{=}\,0.05$ and $\tan\!\beta\,{=}\,2,\,1,\,0.5,\,0.2$ define $\beta$-solitons with $\lambda_{\beta}\,{=}\,5d,\,6.3d,\,10d,\,22.8d$, respectively. 

\begin{figure*}
 \begin{center}
 \includegraphics[width=0.8\textwidth]{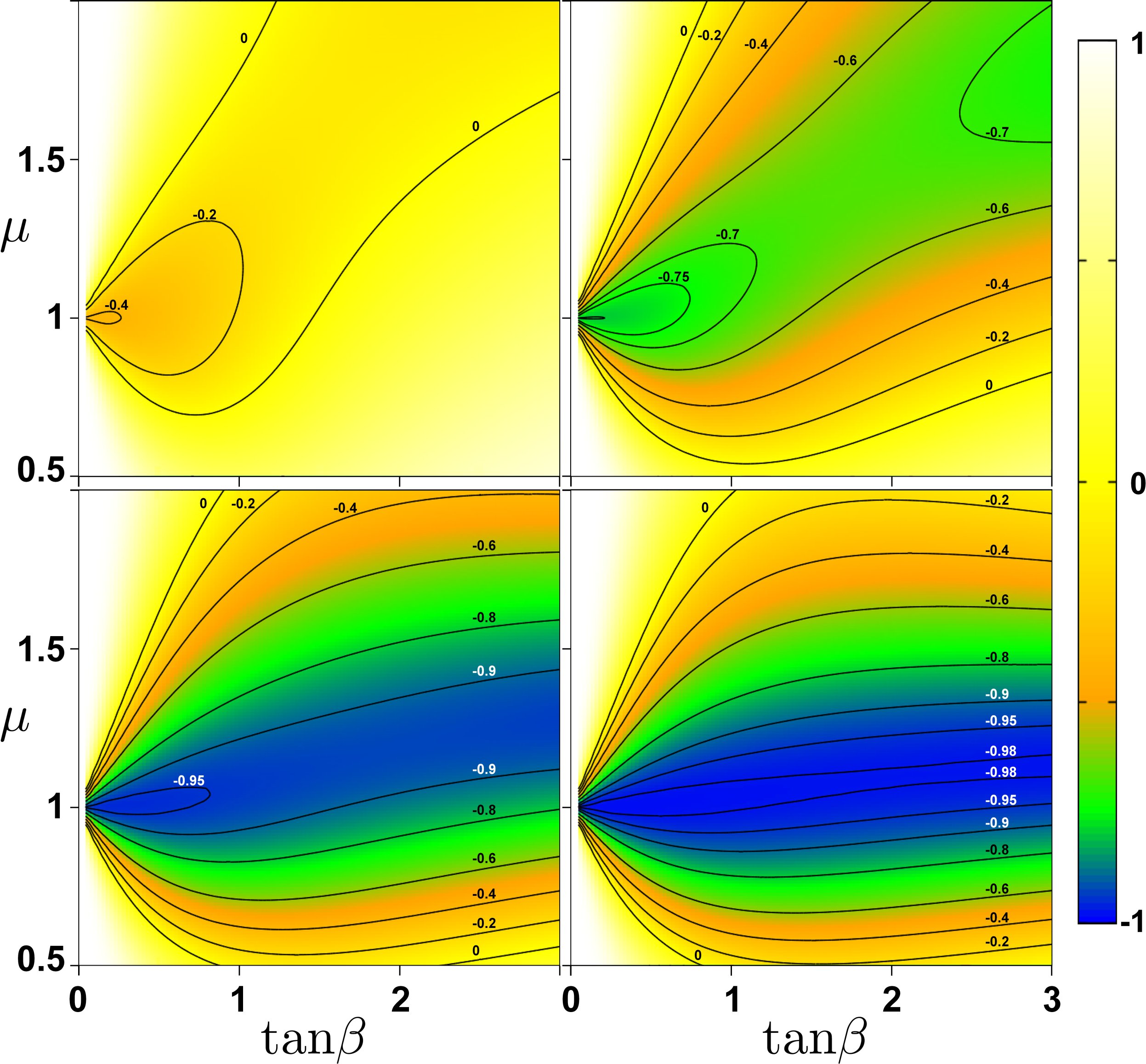}
 \end{center}
 \caption{Contour plots of the asymptotic magnetization $n^z(\tau_{\rm f})$ as a function of the parameters $\tan\!\beta$ and $\mu$, for $\alpha\,{=}\,3$ and $\delta\,{=}\,1$. In each panel a different value of $h$ is considered.}
\label{f.alpha3}
\end{figure*}

In Fig.~\ref{f.alpha3} we show a contour-plot relative to $n^z(\tau_{\rm{f}})$ in the plane ($\tan\!\beta$,$\mu$), for $\delta\,{=}\,1$ and different values of $h$. As expected, for smaller $\tan\!\beta$ the qubit's dynamics is less affected by the finite interaction range, as broad solitons ($\lambda_\beta\gg\alpha{d}$) are little modified by the `smearing' entailed by Eq.~\eqref{e.stildedisc}. In particular, the plot for $h=0.02$ shows that the partial flip shown in Fig.~\ref{f.lapseopt} can be improved by taking smaller $\beta$, i.e. longer solitons, or increasing $\mu$, i.e. the qubit-chain coupling. The flip quality decreases also when, due to the phase term $\xi/\cot\!\beta$ appearing in Eq.~\eqref{e.TWsol}, the $x$ and $y$ components of $\tilde{\bm s}(t)$ shrink under the smearing ~\eqref{e.stildedisc}. This effect can be reduced, as suggested by Eq.~\eqref{e.xidiscr}, by requiring $\alpha\sqrt{h}\ll\cos\!\beta$, i.e., for small $\beta$, $h\ll\alpha^{-2}$. In fact, Fig.~\ref{f.alpha3} shows that by taking a smaller $h$ the flip quality can be made to approach optimal values in an extended region of the $\beta$-$\mu$ plane. Referring to the definitions~\eqref{e.deltamu}, this optimization can be typically performed by driving the external field only.

\subsection{qubit flipped by a generated soliton}
\label{ss.injected}

\begin{figure}
 \includegraphics[width=0.48\textwidth]{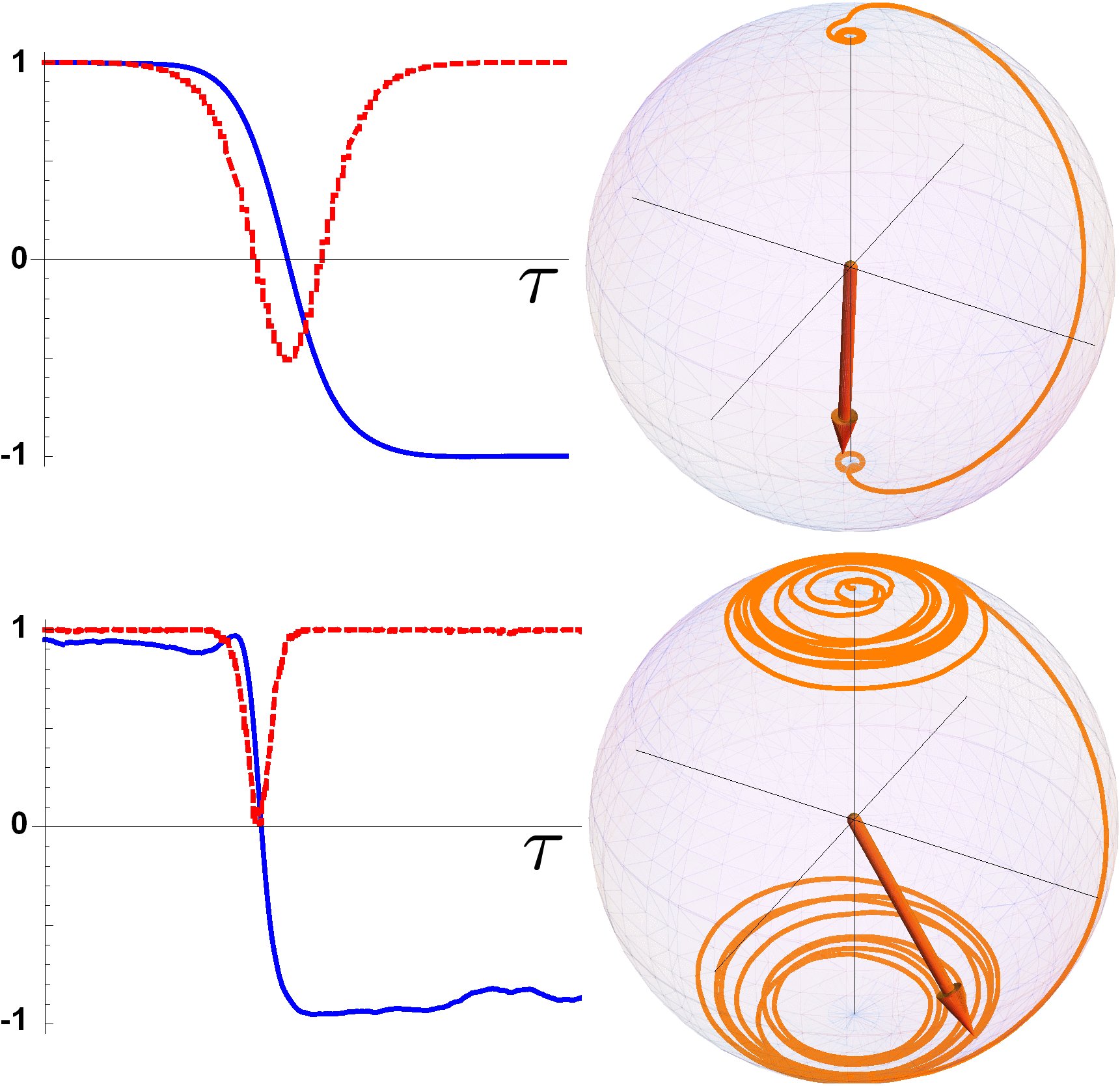}
 \caption{Left panels: time evolution of $n^z(\tau)$ (solid blue line) for the qubit interacting with the soliton ($z$-component, dashed red line) generated by injecting a soliton  ($\tan\!\beta\,{=}\,2$) with ${\cal T}=0$ (top) and ${\cal T}=0.002$ (bottom); $\mu\,{=}\,1$, $\delta\,{=}\,1$, and $\alpha\,{=}\,0$. Right panels: parametric plots of the qubit's state evolution on the Bloch sphere, under the same conditions of the respective left panel.}
	\label{f.tb20_gen}
\end{figure}

Let us now consider the case when the soliton running through the chain  is not ideal, but rather a generated one, ${\bm \Gamma}_l(\beta;\tau)\simeq{\bm \Sigma}^{(\beta')}_l(\tau)$. In Fig.~\ref{f.tb20_gen} we show the qubit's state evolution when $\tan\!\beta\,{=}\,2$ and $\lambda_\beta\,{=}\,5$ ($\alpha\,{=}\,0$): in the left panels one can appreciate that the evolution of $n^z(\tau)$ follows that of the generated soliton, both for zero (top) and finite (bottom) temperature; the right panels display the overall trajectory of the qubit's magnetization on the Bloch sphere. The qubit's behaviour under the action of a generated soliton looks similar to that described in the previous sections for ideal solitons: in particular, for ${\cal{T}}\,{=}\,0$ an almost complete flipping is obtained. More pronounced differences emerge for ${\cal{T}}\,{\neq}\,0$, where the asymptotic value $n^z(\tau_{\rm f})$ is no more constant in time but fluctuates, being subjected to the thermal fluctuations of the spin chain. However, we note that such fluctuations are conceptually different from the decoherence phenomena commonly met when dealing with open quantum systems, as the (whatever noisy) effective field acting on the qubit is still classical, keeping the qubit evolution \textit{on} the Bloch sphere.

\section{Conclusions}
\label{s.Conclusions}

Using dynamical solitons as magnetic signals running through spin-chains is quite a promising prospect, that however needs an in-depth analysis in order to become a more solid possibility. In fact, besides checking intuitions, quantitative conditions must also emerge and be tested, with specific reference to the realization one has in mind. To this respect, the scheme presented in this work might find several different applications, as spin chains are versatile models that can be used for describing the most diverse real situations. Let us therefore end this work by briefly commenting upon the conditions identifiable as essential in our analysis, in the case of an implementation based on solid-state systems~\cite{deLangeWRDH10,BalasubramanianEtal2009,WedgeEtal2012,KoppensBTVNMKV2006,PlaTDLMZJDM2013}.

First of all we have numerically demonstrated that systems of interacting magnetic moments on one-dimensional lattices, possibly of finite length, support dynamical configurations which are the discrete counterparts of $\beta$-solitons if $\lambda_\beta\,{\gg}\,d$, i.e., $\sqrt{\gamma{H}/JS}\sin\!\beta\ll{1}$, to be confronted with $H\neq 0$ for getting $v>0$. Given the values of $S$ and $J$ tipically observed in magnetic compounds, it is $JS^2\,{\sim}\,1\div10^3\,$K meaning that, as $\mu_{\rm{B}}=0.67$~K/Tesla, only very large fields could break the above inequality, and the continuum approximation is therefore most often justified. Further notice that our scheme might also be considered in the case of dynamical solitons that do not require $H\neq 0$ for being supported by the Heisenberg chain~\cite{TjonW1977}. Moreover, we know that solitons exist and run also in anisotropic spin chains~\cite{LongB1979,KosevichIK1990}, which makes our scheme potentially efficient in the case of anisotropic quasi one-dimensional real compounds. In fact, this is quite a relevant feature when thinking of implementations based on one-dimensional monatomic metal chains deposited on surfaces~\cite{GambardellaDMMEKC2002} where the system's geometry inevitably makes the intra-chain exchange anisotropic~\cite{VindigniRPCG2005}.

As for the issue of how dynamical solitons can be generated, we have preliminary indications that our proposal keeps being effective when the action of the field pulse ${\bm b}^{(\beta)}(t)$ is not punctual, as understood by Eq.~\eqref{e.lefttail}, but rather extends to a finite part of the chain end (shorter than the induced soliton), as required in a realistic setup; hence, the effect of ${\bm b}^{(\beta)}(t)$ is amplified by the effective number of spins it affects and its intensity can well be comparable with $H$. 

Having shown that a controlled action on the qubit can actually be obtained by its interaction with the nearby running magnetic soliton, we notice that the condition required by the continuum approximation is fully consistent with the small values of $\sqrt{\gamma H/JS}$ that are found to produce a permanent variation of the qubit state, according to the analysis presented in Sec.~\ref{s.qubit-dynamics}. Moreover, the energy exchanged between qubit and chain in the case of complete flipping (obtained by, say, $\gamma H/JS\,{=}\,0.05$, $\alpha\,{=}\,0$, $\tan\!\beta\,{=}\,0.2$, $\mu\,{=}\,\delta\,{=}\,1$) amounts to $\delta{E}=\hbar(gS{+}\gamma_\sigma H)\,{\simeq}\,10^{-2}JS^2$; as the soliton energy is of the order of $JS^2$, the chain dynamics results unaffected by the evolution of the qubit, essentially validating the `no back-action' approximation mentioned in Sec.~\ref{s.qubit-dynamics}. As for the limits dictated by the typical coherence times attainable in  solid state qubits realizations, an additional relevant quantity is the time $t_{\rm prop}$ required by the soliton to reach the qubit after its injection: for a time scale of $(JS)^{-1}\sim 10^{-13}\,$s, we can estimate $t_{\rm prop}\sim 1\,$ns, if the qubit lies around $10^3$ lattice constants away from the chain end. 

Finally, it is worth noticing that the magnetic solitons propagation we have studied is an energy-conservative phenomenon and has proved to be robust against thermal noise up to a reduced temperature ${\cal T}\sim 0.01$: this suggests that, besides the specific proposal presented in this work, using solitons for transferring either classical or quantum information in solid-state devices might strongly alleviate the heat dissipation requirements that seriously affect more conventional solutions, without requiring a highly demanding lowering of the operating temperature.

\acknowledgments
This work is done in the framework of the {\em Convenzione operativa} between  the Institute for Complex Systems of the Italian National  Research Council, and the Physics and Astronomy Department of the  University of Florence.


\begin{thebibliography}{10}
	
	\bibitem{TejadaCBHS2001}
	J.~Tejada, E.~M.~Chudnovsky, E.~del Barco, J.~M.~Hernandez, and P.~T.~Spiller,
%	Magnetic qubits as hardware for quantum computers
	Nanotechnology {\bf 12}, 181 (2001).
	
	\bibitem{Engeletal2004}
	H.-A.~Engel, L.~P.~Kouwenhoven, D.~Loss, and C.~M.~Marcus,
%	Controlling spin qubits in quantum dots
	Quantum Information Processing {\bf 3}, 115 (2004).
	
	\bibitem{Gaebeletal2006}
	T.~Gaebel, M.~Domhan, I.~Popa, C.~Wittmann, P.~Neumann, F.~Jelezko, J.-R.~Rabeau, N.~Stavrias, A.~D.~Greentree, S.~Prawer, J.~Meijer, J.~Twamley, P.~R.~Hemmer, and J.~Wrachtrup,
%	Room-temperature coherent coupling of single spins in diamond
	Nature Physics {\bf 2}, 408 (2006).
	
	\bibitem{Mazeetal2008}
	J.~R.~Maze, P.~L.~Stanwix, J.~S.~Hodges, S.~Hong, J.~M.~Taylor, P.~Cappellaro, L.~Jiang, M.~V.~Dutt Gurudev, E.~Togan, A.~S.~Zibrov, A.~Yacoby, R.~L.~Walsworth, and M.~D.~Lukin,
%	Nanoscale magnetic sensing with an individual electronic spin in diamond,
	Nature {\bf 455}, 644 (2008).
	
	\bibitem{Auerbach1994}
	A.~Auerbach,
	{\em Interacting electrons and quantum magnetism}
	(Springer, Berlin, 1994)
	
	\bibitem{LongB1979}
	K.~A.~Long and A.~R.~Bishop,
%	Nonlinear excitations in classical ferromagnetic chains
	J. Phys. A {\bf 12} 1325 (1979).
	
	\bibitem{Fogedby1980}
	 H.~C.~Fogedby,
 %	Solitons and magnons in the classical Heisenberg chain 
	 Journal of Physics A, {\bf 13}, 1467 (1980).

	\bibitem{ElstnerM1989}
	N.~Elstner and H.-J.~Mikeska,
%	Solitons in the anisotropic xy chain: semiclassical treatment of quantum effects 
	Journal of Physics: Condensed Matter {\bf 1}, 1487 (1989).
	
	\bibitem{KosevichIK1990}
	A.~M.~Kosevich, B.~A.~Ivanov, and A.~S.~Kovalev,
%	Magnetic solitons
	Physics Reports {\bf 194}, 117 (1990).
	
	\bibitem{SchmidtSL2011}
	H.-J.~Schmidt, C.~Schr\"oder, and M.~Luban,
%	Modulated spin waves and robust quasi-solitons in classical Heisenberg rings
	Journal of Physics: Condensed Matter {\bf 23}, 386003 (2011).
	
	\bibitem{WoellertH2012}
	A.~W\"ollert and A.~Honecker,
%	Solitary excitations in one-dimensional spin chains
	Physical Review B {\bf 85}, 184433 (2012).
	
	\bibitem{KjemsS1978}
	J.~K.~Kjems and M.~Steiner
%	Evidence for soliton modes in the one-dimensional ferromagnet CsNiF$_3$
	Physical Review Letters {\bf 41}, 1137 (1978).
	
	\bibitem{BorsaPRT1983}
	F.~Borsa, M.~G.~Pini, A.~Rettori, and V.~Tognetti
%	Magnetic specific-heat contributions from linear {\em vis-\`a-vis} nonlinear excitations in the one-dimensional antiferromagnet (CH$_3$)$_4$NMnCl$_3$ (TMMC)
	Physical Review B {\bf 28}, 5173 (1983).
	
	\bibitem{PiniR1984}
	M.~G.~Pini and A.~Rettori,
%	Failure of the classical approximation for CsNiF$_3$
	Physical Review B {\bf 29}, 5246 (1984).
	
	\bibitem{JensenMFHS1985}
	H.~J.~Jensen, O.~G.~Mouritsen, H.~C.~Fogedby, P.~Hedeg{\aa}rd and A.~Svane,
%	Analytical and numerical studies of the easy-plane antiferromagnetic chain: Application to (CH$_3$)$_4$NMnCl$_3$
	Physical Review B {\bf 32}, 3240 (1985).
	
	\bibitem{JohnsonW1985}
	M.~D.~Johnson and N.~F.~Wright,
%	Soliton specific heat of spin chains: Limitations of the quantum sine-Gordon model
	Physical Review B {\bf 32}, 5798 (1985).
	
	\bibitem{MikeskaF1986}
	H.-J.~Mikeska and H.~Frahm,
%	The soliton contribution to the specific heat of CsNiF$_3$: quantum effects and out-of-plane fluctuations
	Journal of Physics C: Solid State Physics {\bf 19}, 3203 (1986).
	
	\bibitem{CTVV1991}
	A.~Cuccoli, V.~Tognetti, P.~Verrucchi, and R.~Vaia,
%	Quantum thermodynamics of easy-plane ferromagnetic chains
	Physical Review B {\bf 44}, 903 (1991).
	
	\bibitem{CTVV1992}
	A.~Cuccoli, V.~Tognetti, R.~Vaia, and P.~Verrucchi,
%	Quantum thermodynamics of the easy-plane ferromagnetic chain
	Physical Review B {\bf 46}, 11601 (1992).
	
	\bibitem{CGTVV1995}
	A.~Cuccoli, R.~Giachetti, V.~Tognetti, R.~Vaia, and P.~Verrucchi,
%	The effective potential and effective Hamiltonian in quantum statistical mechanics
	Journal of Physics: Condensed Matter {\bf 7}, 7891 (1995).
	
	\bibitem{CTVV2000}
	A.~Cuccoli, V.~Tognetti, P.~Verrucchi, and R.~Vaia,
%	Semiclassical approach to the thermodynamics of spin chains
	Physical Review B {\bf 62}, 57 (2000).
	
	\bibitem{BoucherRRRBS1980}
	J.~P.~Boucher, L.~P.~Regnault, J.~Rossat-Mignod, J.~P.~Renard, J.~Bouillot, and W.~G.~Stirling,
%	Solitons in the one—dimensional antiferromagnet TMMC
	Solid State Communications {\bf 33}, 171 (1980).
	
	\bibitem{RamirezW1982}
	A.~P.~Ramirez and W.~P.~Wolf,
%	Specific heat of CsNiF$_3$: Evidence for spin solitons
	Physical Review Letters {\bf 49}, 227 (1982).
	
	\bibitem{MikeskaS1991}
	H.-J.~Mikeska and M.~Steiner,
%	Solitary excitations in one-dimensional magnets
	Advances in Physics {\bf 40}, 191 (1991).
	
	\bibitem{TjonW1977}
	J.~Tjon and J.~Wright,
%	Solitons in the continuous Heisenberg spin chain
	Physical Review B {\bf 15}, 3470 (1977).
	
	\bibitem{Takhtajan1977}
	L.~A.~Takhtajan,
%	Integration of the continuous Heisenberg spin chain through the inverse scattering method
	Physics Letters A {\bf 64}, 235 (1977).
	
	\bibitem{KrechBL1998}
	M.~Krech, A.~Bunker, and D.P.~Landau,
%	Fast spin dynamics algorithms for classical spin systems
	Computer Physics Communications {\bf 111}, 1 (1998).
	
	\bibitem{Yoshida1990}
	H.~Yoshida,
%	Construction of higher order symplectic integrators
	Physics Letters A {\bf 150}, 262 (1990).
	
	\bibitem{ForestR1990}
	E.~Forest and R.~D.~Ruth
%	Fourth-order symplectic integration
	Physica D {\bf 43}, 105 (1990).
	
	\bibitem{Suzuki1992}
	M.~Suzuki
%	General theory of higher-order decomposition of exponential operators and symplectic integrators
	Physics Letters A {\bf 165}, 387 (1992).
	
	\bibitem{TsaiLL2005}
	S.-H.~Tsai, H.~K.~Lee, and D.~P.~Landau,
%	Molecular and spin dynamics simulations using modern integration methods
	American Journal of Physics {\bf 73}, 615 (2005).
	
	\bibitem{CNVV2014a}
	A.~Cuccoli, D.~Nuzzi, R.~Vaia, and P.~Verrucchi,
%	Quantum gates controlled by spin chain soliton excitations
	Journal Applied Physics {\bf 115}, 17B302 (2014).
	
	\bibitem{CNVV2014b}
	A.~Cuccoli, D.~Nuzzi, R.~Vaia, and P.~Verrucchi,
%	Using solitons for manipulating qubits
	International Journal of Quantum Information {\bf 12}, 1461013 (2014).
	
	\bibitem{deLangeWRDH10}
	G.~de~Lange, Z.~H.~Wang, D.~Rist\`e, V.~V.~Dobrovitski, and R.~Hanson,
%	Universal dynamical decoupling of a single solid-state spin from a spin bath
	Science, {\bf 330} 60 (2010).
	
	\bibitem{BalasubramanianEtal2009}
	G.~Balasubramanian, P.~Neumann, D.~Twitchen, M.~Markham, R.~Kolesov, N.~Mizuochi, J.~Isoya, J.~Achard, J.~Beck, J.~Tissler, V.~Jacques, P.~R.~Hemmer, F.~Jelezko, and J.~Wrachtrup,
%	Ultralong spin coherence time in isotopically engineered diamond
	Nature Materials {\bf 8}, 383 (2009).
	
	\bibitem{WedgeEtal2012}
	C.~J.~Wedge, G.~A.~Timco, E.~T.~Spielberg, R.~E.~George, F.~Tuna F, S.~Rigby, E.~J.~L.~McInnes, R.~E.~P.~Winpenny, S.~J.~Blundell, and A.~Ardavan,
%	Chemical engineering of molecular qubits
	Physical Review Letters {\bf 108}, 107204 (2012).
	
	\bibitem{KoppensBTVNMKV2006}
	F~H~L.~Koppens, C.~Buizert, K.~J.~Tielrooij, I.~T.~Vink, K.~C.~Nowack, T.~Meunier, L.~P.~Kouwenhoven, and L.~M.~K.~Vandersypen,
%	Driven coherent oscillations of a single electron spin in a quantum dot
	Nature {\bf 442}, 766 (2006).
	
	\bibitem{PlaTDLMZJDM2013}
	J.~J.~Pla, K.~Y.~Tan, J.~P.~Dehollain, W.~H.~Lim, J.~J.~L.~Morton, F.~A.~Zwanenburg, D.~N.~Jamieson, A.~S.~Dzurak, and A.~Morello,
%	High-fidelity readout and control of a nuclear spin qubit in silicon
	Nature {\bf 496}, 334 (2013).

	\bibitem{GambardellaDMMEKC2002}
	P.~Gambardella, A.~Dallmeyer, K.~Maiti, M.~C.~Malagoli, W.~Eberhardt, K.~Kern, and C.~Carbone,
%	Ferromagnetism in one-dimensional monatomic metal chains
	Nature {\bf 416}, 301 (2002).
	
	\bibitem{VindigniRPCG2005}
	A.~Vindigni, A.~Rettori, M.~G.~Pini, C.~Carbone, P.~Gambardella,
%	Finite-sized Heisenberg chains and magnetism of one-dimensional metal system
	Applied Physics A {\bf 82}, 385 (2005).
		
\end{thebibliography}
\end{document}